\documentclass[10pt,aps,prd,twocolumn,superscriptaddress,nofootinbib,floatfix]{revtex4-2}
\usepackage{amsmath,amssymb,bm}
\usepackage{graphicx}
\usepackage{booktabs}
\usepackage[colorlinks=true,linkcolor=red,citecolor=red,urlcolor=magenta]{hyperref}
\usepackage{orcidlink}

\begin{document}

\title{Kalb-Ramond black bounces: throat-blind shadows and throat-sensitive neutrinos}

\author{Faizuddin Ahmed\orcidlink{0000-0003-2196-9622}}
\email[Email: ]{faizuddinahmed15@gmail.com}
\affiliation{Department of Physics, The Assam Royal Global University, Guwahati 781035, Assam, India}

\author{Ahmad Al-Badawi\orcidlink{0000-0002-3127-3453}}
\email[Email: ]{ahmadbadawi@ahu.edu.jo}
\affiliation{Department of Physics, Al-Hussein Bin Talal University, P. O. Box 20, 71111, Ma'an, Jordan}

\author{\.{I}zzet Sakall\i\orcidlink{0000-0001-7827-9476}}
\email[Email: ]{izzet.sakalli@emu.edu.tr (Corresponding author)}
\affiliation{Physics Department, Eastern Mediterranean University, Famagusta 99628, North Cyprus via Mersin 10, T\"urkiye}

\begin{abstract}
Regularizing a black hole removes its singularity, but which observations could detect the regularized core? We address this question for a Simpson-Visser black bounce built on the power-law hairy black hole that a Lorentz-violating Kalb-Ramond background produces. Measured in the areal radius, the bounce leaves the metric function untouched and changes only the radial measure. This single fact sorts the observables into two classes. The shadow, the innermost stable orbit, orbital frequencies and gravitational redshifts are blind to the throat. The instability of the photon ring, the brightness of its subrings, the oscillation phase of neutrinos and the energy deposited by neutrino pair annihilation respond to it through one universal factor. We derive an exact law for the subring demagnification that holds for any metric function. We also show that the throat enters the weak-field neutrino phase and the light deflection with different weights, so that combined photon and neutrino lensing can separate the mass from the throat. Ray-traced polarized images of a turbulent thin disk confirm that the direct image hides the throat while the first subring reveals it. For neutrinos emitted near a stellar-mass compact object the throat reshapes the flavor survival pattern and enhances the annihilation energy deposition. Two independent computations reproduce each headline number.\\

{\bf keywords:} Modified gravity theory; black bounce; geodesic motions; neutrino phase; 
\end{abstract}

\maketitle

\section{Introduction}\label{isec1}

The images of M87* and Sagittarius A* released by the Event Horizon Telescope (EHT) have turned the size of the black-hole critical curve into a measured quantity \cite{EHT2019,EHT2022}, and the polarimetric maps of M87* have added the pattern of the electric vector position angle around the ring \cite{EHT2021}. These data test the metric function that controls circular photon orbits. They are much less sensitive to the structure of the spacetime inside the photon sphere, and this is precisely where proposals for singularity resolution put their new physics. The Simpson-Visser construction is the cleanest example \cite{SimpsonVisser2019}: replacing the areal radius $r$ by $\sqrt{r^2+a^2}$ turns the Schwarzschild singularity into a throat of radius $a$ while leaving the shadow of the black hole untouched. Lensing \cite{Nascimento2020,Tsukamoto2021a,Tsukamoto2021b,Islam2021,He2024}, thin-disk images \cite{Guerrero2021} and geodesic completeness \cite{Pal2023} of such black bounces have been studied in detail, and their charged version has a known Reissner-Nordstr\"om limit \cite{Franzin2021}. Their field sources in general relativity require a phantom scalar together with a nonlinear electromagnetic field \cite{Bronnikov2022}.

A second line of work concerns spontaneous Lorentz symmetry breaking in the gravitational sector. The antisymmetric Kalb-Ramond field \cite{KalbRamond1974}, with a nonzero vacuum expectation value coupled nonminimally to the Ricci tensor, is a natural candidate within the Standard-Model Extension \cite{Altschul2010}. Lessa and collaborators found that a radial pseudo-electric expectation value produces an exact power-law hair, $f(r)=1-2M/r+\Upsilon r^{-2/\lambda}$, controlled by the product $\lambda$ of the coupling and the norm of the background \cite{Lessa2020}. Other choices of the Lagrange multiplier lead to different families \cite{Yang2023}, and rotating and charged versions have been used for shadows and photon dynamics \cite{Kumar2020,Sucu2025}. Lorentz-violating gravity also admits regular and geodesically complete black holes of its own \cite{CarballoRubio2022,Mazza2023}.

Neutrinos are the second messenger that can reach the inner region. Their oscillation phase accumulated along a curved null path was derived in the Schwarzschild metric long ago \cite{Fornengo1997,Cardall1997}, generalized to arbitrary static backgrounds \cite{Visinelli2015}, and extended to spin precession \cite{Bruggen1998}. When a lens creates two or more paths, the flavor probability acquires interference terms sensitive to the absolute neutrino masses and to the mass ordering \cite{Swami2020}, with decoherence effects analyzed in Ref.~\cite{Swami2021}. This program has since been carried through a long list of geometries \cite{Capozziello2000,Chakrabarty2022,Chakrabarty2023,Alloqulov2025,Mannobova2026}, and the possibility of lensed astrophysical neutrinos has been raised observationally \cite{Britzen2026}. The pair-annihilation process $\nu\bar\nu\to e^+e^-$ above neutrinospheres and accretion disks offers a third handle, since its energy deposition is strongly amplified by gravity \cite{Asano2000,Asano2001,Birkl2007} and it is sensitive to modified gravity \cite{Lambiase2020,Lambiase2021,Lambiase2022}. Most directly relevant here, Shi and Ara\'ujo Filho studied annihilation, oscillation phases and lensed flavor probabilities in the Kalb-Ramond black hole \cite{Shi2025KR,Shi2026KR}, in bumblebee gravity \cite{Shi2025BB,Shi2026NM}, and Lorentz-breaking corrections to the annihilation jets of gamma-ray bursts were discussed in Ref.~\cite{Khodadi2023}. The Simpson-Visser metric has been confronted with the deposition rate as well \cite{AraujoFilho2024SV}.

The two ingredients are complementary. The Kalb-Ramond hair is an infrared modification controlled by a background tensor that pervades the exterior, while the throat is an ultraviolet modification of the core that leaves the exterior metric function intact. A geometry carrying both is the natural testing ground for a question that neither ingredient raises alone: which observables respond to the core and which only to the exterior, and can a single observing campaign tell them apart? Photons answer part of this question through the photon ring, whose subrings are governed by the instability of the bound photon orbit \cite{Johnson2020,Gralla2019,Cardoso2009} and whose polarization follows a pattern fixed by the near-ring geometry \cite{Himwich2020,Narayan2021,Palumbo2020}. Neutrinos answer another part, because the oscillation phase is an integral over the whole path rather than a property of a single orbit.

These studies treat each observable one at a time. None of them separates the information carried by $g_{tt}$ as a function of the areal radius from the information carried by the radial measure, yet for a Simpson-Visser regularization the first is exactly unchanged and the second carries all the dependence on the throat. Whether neutrino phases and annihilation can see a throat that the shadow cannot, and how that compares with what the photon subrings and polarization can do, has therefore not been tested.

This paper fills that gap for the Kalb-Ramond black bounce. Three technical elements make the comparison possible. We show that in any static spherically symmetric spacetime the polarization of a ray can be transported with two conserved components, without a Walker-Penrose constant, so that polarized images of the bounce are computed exactly. We weight neutrino paths by the ratio of emitted to received solid angle rather than by equal amplitudes, which matters once paths wind around the photon sphere. And we identify the proper volume element as the only route by which the throat enters the annihilation rate. Section~\ref{isec2} sets up the field theory, the regularized metric and its effective source. Section~\ref{isec3} proves the split between throat-blind and throat-sensitive observables and derives the weak-field coefficients, including the combination of photon deflection and neutrino phase that separates the mass from the throat. Section~\ref{isec4} presents polarized ray-traced images and their subrings, Sec.~\ref{isec5} the neutrino phases and multipath flavor probabilities, and Sec.~\ref{isec6} the annihilation deposition. Section~\ref{isec7} translates the results into observational windows, and Sec.~\ref{isec8} concludes.

\section{Kalb-Ramond hair and its Simpson-Visser regularization}\label{isec2}

We use the signature $(-,+,+,+)$ and units $G=c=1$; in all numerical work lengths are measured in units of the mass $M$. The construction has two steps, the derivation of the hairy solution in the Kalb-Ramond theory and the regularization of its center, and we keep them apart because the second step is not a solution of the field equations of the first.

\subsection{Field equations and the power-law hair}

The Kalb-Ramond field is a two-form $B_2=\frac12 B_{\mu\nu}dx^\mu\wedge dx^\nu$ with field strength $H_{\lambda\mu\nu}=\partial_{[\lambda}B_{\mu\nu]}$ \cite{KalbRamond1974}. A self-interaction $V(B_{\mu\nu}B^{\mu\nu}\pm b_{\mu\nu}b^{\mu\nu})$ with a nonzero vacuum value $\langle B_{\mu\nu}\rangle=b_{\mu\nu}$ breaks local Lorentz symmetry spontaneously \cite{Altschul2010}. With a nonminimal coupling to curvature the action reads \cite{Lessa2020}
\begin{align}
S=\int d^4x\sqrt{-g}\Big[&\frac{R}{2\kappa}-\frac{1}{12}H_{\lambda\mu\nu}H^{\lambda\mu\nu}-V\nonumber\\
&+\frac{1}{2\kappa}\left(\xi_2B^{\lambda\nu}B^{\mu}{}_{\nu}R_{\lambda\mu}+\xi_3B^{\mu\nu}B_{\mu\nu}R\right)\Big],\label{eq:action}
\end{align}
with $\kappa=8\pi G$ and couplings $\xi_{2,3}$ of dimension length squared. On the vacuum $B_{\mu\nu}B^{\mu\nu}=b_{\mu\nu}b^{\mu\nu}=b^2$ the $\xi_3$ term is a constant rescaling of $R$ and drops out. For a static spherically symmetric metric $ds^2=-A\,dt^2+B\,dr^2+r^2d\Omega^2$ the pseudo-electric configuration
\begin{equation}
b_2=-\tilde E(r)\,dt\wedge dr,\qquad \tilde E(r)=|b|\sqrt{\frac{A(r)B(r)}{2}},\label{eq:vev}
\end{equation}
has constant norm and vanishing field strength. Varying Eq.~\eqref{eq:action} with respect to the metric gives $G_{\mu\nu}=\kappa T^{\xi_2}_{\mu\nu}$, whose $tt$ and $rr$ components reduce to $(1-\lambda/2)R_{tt}=(1-\lambda/2)R_{rr}=0$, where
\begin{equation}
\lambda\equiv\xi_2|b|^2 .
\end{equation}
For $\lambda\neq2$ these give $A=1/B$, and the angular equation becomes
\begin{equation}
\frac{\lambda}{2}r^2A''+(\lambda+1)rA'+A-1=0 .\label{eq:ode}
\end{equation}
The trial power $A-1\propto r^p$ yields $(\lambda p/2+1)(p+1)=0$, so the general solution is \cite{Lessa2020}
\begin{equation}
A(r)=1-\frac{2M}{r}+\frac{\Upsilon}{r^{n}},\qquad n\equiv\frac{2}{\lambda},\label{eq:lessa}
\end{equation}
where $\Upsilon$ has dimension length$^{n}$. For $\lambda=1$ the hair has the Reissner-Nordstr\"om fall-off with $\Upsilon$ in place of $Q^2$, and for $\lambda=-1$ it mimics a cosmological constant. We adopt $\lambda=2/3$, so $n=3$, as the fiducial value: the hair then decays faster than any electromagnetic charge and the solution is not a reparametrization of a known charged geometry.

\subsection{Regularization and horizon structure}

The metric \eqref{eq:lessa} has a curvature singularity at $r=0$. Following Simpson and Visser \cite{SimpsonVisser2019}, we replace the areal radius by $R(r)=\sqrt{r^2+a^2}$ everywhere,
\begin{align}
ds^2&=-f(R)\,dt^2+\frac{dr^2}{f(R)}+R^2\,d\Omega^2,\nonumber\\
f(R)&=1-\frac{2M}{R}+\frac{\Upsilon}{R^{n}},\label{eq:svkr}
\end{align}
with $r\in(-\infty,\infty)$ and a throat at $r=0$ where $R=a$. Since $R\geq a>0$, the metric function and all its derivatives remain bounded and every curvature invariant is finite on the throat. We call the line element \eqref{eq:svkr} the SV-KR geometry. Noted that $\lambda=1$ serves as a cross-check against the charged-like black bounce \cite{Franzin2021}. Changing the radial coordinate to $R$ gives the form that organizes the rest of the paper,
\begin{equation}
ds^2=-f(R)\,dt^2+\frac{dR^2}{h(R)f(R)}+R^2d\Omega^2,\quad h(R)=1-\frac{a^2}{R^2}.\label{eq:areal}
\end{equation}
The metric function in terms of the areal radius is identical to the Kalb-Ramond one, and the bounce appears only through $h$.

\begin{figure*}[t]
\centering
\includegraphics[width=0.94\textwidth]{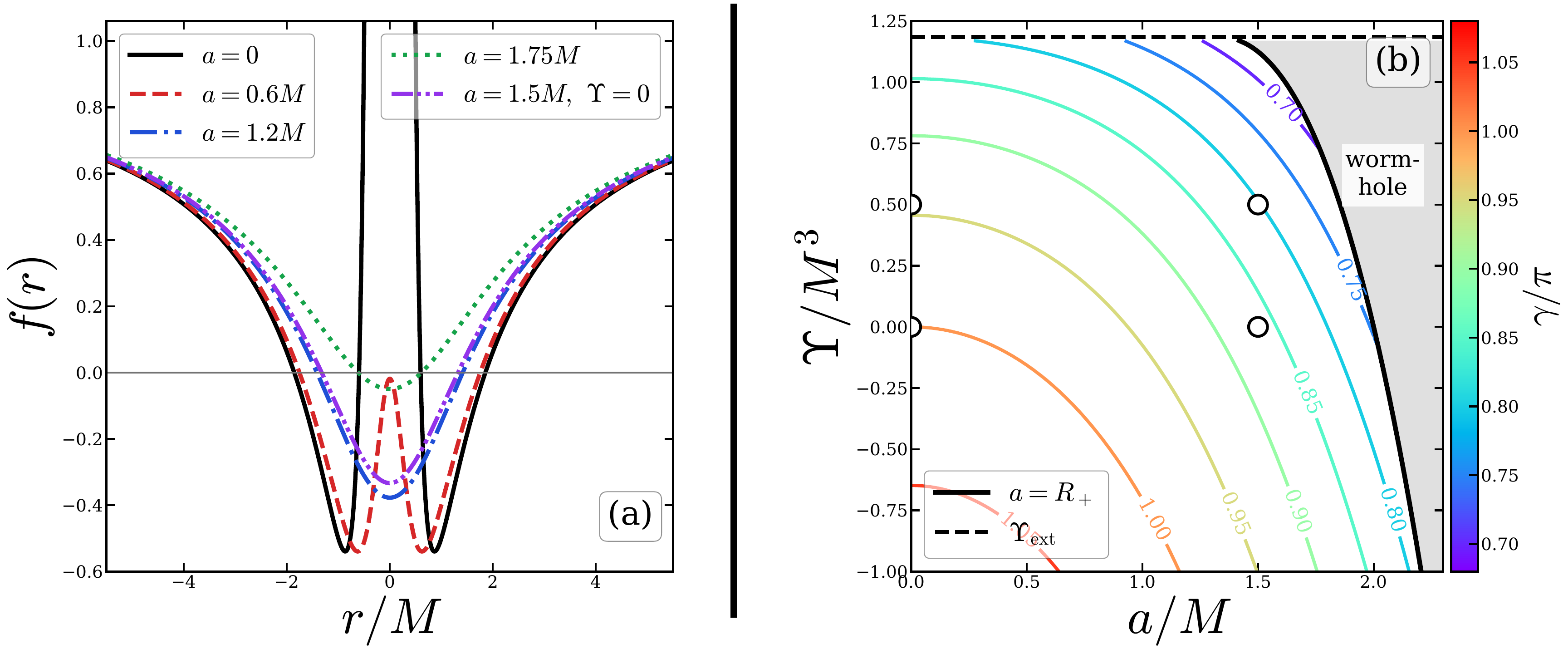}
\caption{(a) Metric function of the SV-KR geometry with $\lambda=2/3$ across the throat $r=0$ for $\Upsilon=0.5M^3$ and increasing throat radius, together with the hairless bounce of radius $1.5M$; the unregularized solution diverges at the center, the regularized ones stay finite. (b) Contours of the subring demagnification exponent $\gamma/\pi$ in the plane of throat radius and hair, each labeled by its value. The hatched region is the wormhole branch, the solid curve $a=R_+$ separates black bounces from the wormhole branch and the dashed line marks the extremal hair $\Upsilon_{\rm ext}=32M^3/27$. Circles mark the four geometries used for the images.}
\label{fig:geo}
\end{figure*}

Horizons are the roots $f(R_\pm)=0$ with $R_\pm>a$. For $\Upsilon>0$ the hair produces an inner horizon, and the two merge at the extremal point $f=f'=0$, which gives
\begin{equation}
R_{\rm ext}=\frac{2(n-1)}{n}M,\qquad \Upsilon_{\rm ext}=\frac{2}{n}R_{\rm ext}^{\,n-1}M ,\label{eq:ext}
\end{equation}
that is, $\Upsilon_{\rm ext}=32M^3/27$ for $n=3$ and $\Upsilon_{\rm ext}=M^2$ for $n=2$. For $\Upsilon=M^3$ and $n=3$ the horizons are $R_-=M$ and $R_+=(1+\sqrt5)M/2$, because $R^3-2MR^2+M^3$ factorizes. The geometry is a black bounce when $a<R_+$ and a wormhole otherwise, and we restrict the analysis to the black-bounce branch. Figure~\ref{fig:geo}(a) shows the metric function across the throat: for $a=0$ the hair makes $f$ diverge at the center, while any $a>0$ turns it into a smooth even function of $r$.

\subsection{Effective source and energy conditions}

The regularized metric is not a vacuum solution of the Kalb-Ramond equations, and it is useful to see exactly what it requires. Reading $\kappa\rho=-G^t{}_t$, $\kappa p_r=G^r{}_r$ and $\kappa p_t=G^\theta{}_\theta$ from Eq.~\eqref{eq:svkr}, a direct computation gives
\begin{align}
\kappa\rho&=\frac{1}{R^{n+5}}\Big[(n-1)\Upsilon R^3+4Ma^2R^{n}\nonumber\\
&\qquad-a^2R^{n+1}-(n+1)\Upsilon a^2R\Big],\label{eq:rho}\\
\kappa p_r&=\frac{-(n-1)\Upsilon R^2-a^2R^{n}+(n-1)\Upsilon a^2}{R^{n+4}},\label{eq:pr}\\
\kappa p_t&=\frac{1}{2R^{n+5}}\Big[n(n-1)\Upsilon R^3-2Ma^2R^n\nonumber\\
&\qquad+2a^2R^{n+1}-(n^2-2)\Upsilon a^2R\Big].\label{eq:pt}
\end{align}
For $a=0$ these reduce to $\kappa\rho=-\kappa p_r=(n-1)\Upsilon/R^{n+2}$, the effective Kalb-Ramond source of the hair \cite{Lessa2020}. The combination that controls the null energy condition takes a simple form,
\begin{equation}
\rho+p_r=-\frac{2a^2f(R)}{\kappa R^4},\label{eq:nec}
\end{equation}
independent of the hair except through $f$. The null energy condition is violated everywhere outside the horizon and its violation is set by the throat alone, exactly as in the hairless bounce. A field theory producing Eq.~\eqref{eq:svkr} must therefore combine the Kalb-Ramond sector with a phantom component of the type found for the Simpson-Visser geometry in general relativity \cite{Bronnikov2022}. In what follows we treat Eq.~\eqref{eq:svkr} as an effective geometry and study its test-particle and test-field content.

Two further properties of the regularized geometry are worth recording. First, since $-g_{tt}g_{rr}=1$ in the coordinate $r$, the vacuum value of Eq.~\eqref{eq:vev} keeps the constant magnitude $\tilde E=|b|/\sqrt2$ along the whole bounce, so the background tensor is smooth through the throat, whereas in the areal coordinate its component grows as $h^{-1/2}$ toward $R=a$. Second, the curvature at the throat is finite. For $n=3$ the Kretschmann scalar at $r=0$ is
\begin{align}
K\big|_{r=0}=\frac{1}{a^{10}}\Big(&12a^6-32Ma^5+36M^2a^4\nonumber\\
&+16\Upsilon a^3-44M\Upsilon a^2+17\Upsilon^2\Big),\label{eq:kretsch}
\end{align}
which for $\Upsilon=0$ becomes $(12a^2-32Ma+36M^2)/a^6$ and diverges only in the singular limit $a\to0$. The radial coordinate $r$ covers two asymptotically flat regions joined through the throat; in the black-bounce branch both sides are separated from the throat by horizons, and all observers considered below sit in the $r>0$ exterior.

\section{Throat-blind and throat-sensitive observables}\label{isec3}

The form \eqref{eq:areal} implies a general statement that holds for any metric function $f(R)$, and the Kalb-Ramond hair is only one realization of it. For geodesics with conserved energy $E$ and angular momentum $L$ in the equatorial plane,
\begin{equation}
\Big(\frac{dR}{d\tau}\Big)^2=h(R)\,f(R)\Big[\frac{E^2}{f}-\epsilon-\frac{L^2}{R^2}\Big],\label{eq:radial}
\end{equation}
with $\epsilon=0$ for null and $\epsilon=1$ for timelike geodesics. The throat multiplies the radial equation by a positive function and does nothing else. Any condition stated as the vanishing of the bracket and of its derivatives is therefore unchanged, while any quantity that integrates along $R$ or depends on the curvature of the radial potential at a fixed point picks up $h$.

\subsection{Throat-blind quantities}

The photon sphere solves $d(f/R^2)/dR=0$, so it satisfies $2f(R_{\rm ph})=R_{\rm ph}f'(R_{\rm ph})$ and the critical impact parameter is
\begin{equation}
b_c=\frac{R_{\rm ph}}{\sqrt{f(R_{\rm ph})}} .\label{eq:bc}
\end{equation}
Timelike circular orbits have
\begin{equation}
E^2=\frac{2f^2}{2f-Rf'},\quad L^2=\frac{R^3f'}{2f-Rf'},\quad \Omega_\phi=\sqrt{\frac{f'}{2R}},\label{eq:circ}
\end{equation}
and the marginally stable orbit solves
\begin{equation}
Rff''-2Rf'^2+3ff'=0 .\label{eq:isco}
\end{equation}
Here prime denotes ordinary derivative w. r. to argument. Since $h$ multiplies the whole bracket of Eq.~\eqref{eq:radial}, the second derivative of the radial potential at a circular orbit only rescales, so the condition for the ISCO is also unchanged. The same holds for the redshift of a static emitter, $\sqrt{f}$, and for the redshift of a circular emitter, $\sqrt{f-R^2\Omega_\phi^2}$. The photon sphere, the shadow, the ISCO, the Keplerian frequency, the radiative efficiency $1-E_{\rm ISCO}$ and all local redshifts are throat-blind, provided the corresponding radius lies above $a$. For the four fiducial geometries Table~\ref{tab:geo} lists $b_c$ and $R_{\rm ISCO}$, which depend on the hair but are the same for $a=0$ and $a=1.5M$.

\subsection{Throat-sensitive quantities}

Expanding Eq.~\eqref{eq:radial} around the photon sphere gives the Lyapunov exponent of the unstable orbit in coordinate time \cite{Cardoso2009},
\begin{equation}
\lambda_L=\sqrt{h(R_{\rm ph})}\,\lambda_L^{(0)},\quad \lambda_L^{(0)}=\sqrt{\frac{f_{\rm ph}\left(2f_{\rm ph}-R_{\rm ph}^2f''_{\rm ph}\right)}{2R_{\rm ph}^2}},\label{eq:lyap}
\end{equation}
where $\lambda_L^{(0)}$ is the value for $a=0$. Successive photon subrings are demagnified by $e^{-\gamma}$ with $\gamma=\pi\lambda_L/\Omega_{\rm ph}$ and $\Omega_{\rm ph}=1/b_c$ \cite{Johnson2020,Gralla2019}. Because $\Omega_{\rm ph}$ is throat-blind, we obtain the exact law
\begin{equation}
\gamma(a)=\gamma(0)\sqrt{1-\frac{a^2}{R_{\rm ph}^2}},\label{eq:gamma}
\end{equation}
valid for any metric function. For the hairless bounce it reads $\gamma=\pi\sqrt{1-a^2/9M^2}$, and for $a=1.5M$ it gives $\lambda_L=1/(6M)$ exactly. The radial epicyclic frequency is likewise rescaled, $\kappa_R^2=h(R)\,\kappa_{R,0}^2$, while the vertical and azimuthal frequencies are untouched; this offers a second, independent probe of $h$ through quasiperiodic oscillations.

Path integrals pick up the factor $h^{-1/2}$. For a null ray of impact parameter $b=L/E$ with turning point $R_0$, the swept angle and the radial integral that enters the neutrino phase are
\begin{align}
\psi(b)&=\int_{R_0}^{\infty}\frac{b\,dR}{R^2\sqrt{h(R)}\sqrt{1-b^2f(R)/R^2}},\label{eq:psi}\\
\mathcal{I}(b;R_{\rm e})&=\int_{R_0}^{R_{\rm e}}\frac{dR}{\sqrt{h(R)}\sqrt{1-b^2f(R)/R^2}} .\label{eq:I}
\end{align}
Expanding for $b\gg M,a$ we find for the deflection angle $\alpha=2\psi-\pi$
\begin{align}
\alpha=&\frac{4M}{b}+\frac{\pi(15M^2+a^2)}{4b^2}+\frac{128M^3}{3b^3}+\frac{8Ma^2}{3b^3}\nonumber\\
&+c^{\alpha}_n\frac{\Upsilon}{b^{n}}+\dots,\label{eq:defl}
\end{align}
and for the regularized radial integral of one leg, $\mathcal{P}(b)\equiv\lim_{R_{\rm e}\to\infty}[\mathcal{I}-R_{\rm e}]$,
\begin{equation}
\mathcal{P}(b)=M+\frac{\pi(3M^2+a^2)}{4b}+c^{\Phi}_n\frac{\Upsilon}{b^{n-1}}+\dots,\label{eq:Pexp}
\end{equation}
with hair coefficients
\begin{equation}
c^{\Phi}_n=-\frac{\sqrt\pi\,\Gamma\!\left(\frac{n+1}{2}\right)}{2\,\Gamma\!\left(\frac n2\right)},\qquad c^{\alpha}_n=\frac{2(n+1)}{n}c^{\Phi}_n .\label{eq:cn}
\end{equation}
For $n=2$ these give $c^\alpha_2=-3\pi/4$, the Reissner-Nordstr\"om deflection coefficient with $\Upsilon=Q^2$, and for $n=3$ they give $c^\Phi_3=-1$ and $c^\alpha_3=-8/3$. The $a^2$ term of Eq.~\eqref{eq:defl} reproduces the Simpson-Visser result of Ref.~\cite{Nascimento2020}. The derivation of Eqs.~\eqref{eq:defl}--\eqref{eq:cn} is given in Appendix~\ref{app:weak}.

The comparison of Eqs.~\eqref{eq:defl} and \eqref{eq:Pexp} is the first quantitative result of this paper. The throat enters both expansions at second order, but with different weights: in the deflection it shifts $M^2\to M^2+a^2/15$, in the two-leg neutrino phase it shifts $M^2\to M^2+a^2/3$. Denoting the two second-order coefficients by $\alpha_2=\pi(15M^2+a^2)/4$ and $\Phi_2=\pi(3M^2+a^2)/2$, one obtains
\begin{equation}
M^2=\frac{2\alpha_2-\Phi_2}{6\pi},\qquad a^2=\frac{5\Phi_2-2\alpha_2}{2\pi},\label{eq:invert}
\end{equation}
so a photon measurement and a neutrino measurement of the same lens separate mass and throat at this order. For $n=3$ the hair enters both coefficients only at higher order, and the inversion is hair-independent. For $n=2$ the hair adds $-3\pi\Upsilon/4$ to $\alpha_2$ and $-\pi\Upsilon/2$ to $\Phi_2$, and Eq.~\eqref{eq:invert} then returns $M^2-\Upsilon/6$ and $a^2-\Upsilon/2$ instead of $M^2$ and $a^2$; the charged black bounce therefore carries a degeneracy between charge and throat that the Kalb-Ramond hair with $\lambda<1$ avoids at this order.

The structure of the problem can be stated in one line. Every observable considered in this paper is either a functional $\mathcal{O}[f]$ of the metric function alone or an integral of the form
\begin{equation}
\mathcal{O}=\int dR\,\frac{\mathcal{W}[f](R)}{\sqrt{1-a^2/R^2}},\label{eq:universal}
\end{equation}
with a kernel $\mathcal{W}$ that again depends on $f$ alone, or a local quantity proportional to $\sqrt{h}$ at a fixed radius, as in Eq.~\eqref{eq:lyap}. The placement of each observable in one of these classes, and the factor $h$ itself, are universal: they hold for any metric function and any throat. The Kalb-Ramond hair is model dependent and enters only through $f$, via the exponent $n=2/\lambda$ and the amplitude $\Upsilon$. This is the sense in which the paper separates a universal statement from a model-dependent one.

For timelike orbits the same logic yields the radial epicyclic frequency in closed form. Perturbing Eq.~\eqref{eq:radial} around a circular orbit and converting to coordinate time,
\begin{equation}
\Omega_R^2=h(R)\,\frac{Rff''-2Rf'^2+3ff'}{2R},\label{eq:epi}
\end{equation}
which vanishes on the ISCO of Eq.~\eqref{eq:isco} and reduces to $M(1-6M/R)/R^3$ for Schwarzschild. The periastron precession per orbit, $2\pi(\Omega_\phi/\Omega_R-1)$, is therefore throat-sensitive, and the ratio of vertical to radial frequencies is shifted by $h^{-1/2}$ at every radius, while the orbital frequency itself is not. For the S2 star around Sagittarius A*, whose pericenter lies near $3\times10^3M$, the factor $h$ differs from unity by about $10^{-7}$ for $a\sim M$, so stellar orbits are effectively throat-blind and the inner disk is where the epicyclic channel would operate.

\begin{figure*}[t]
\centering
\includegraphics[width=0.94\textwidth]{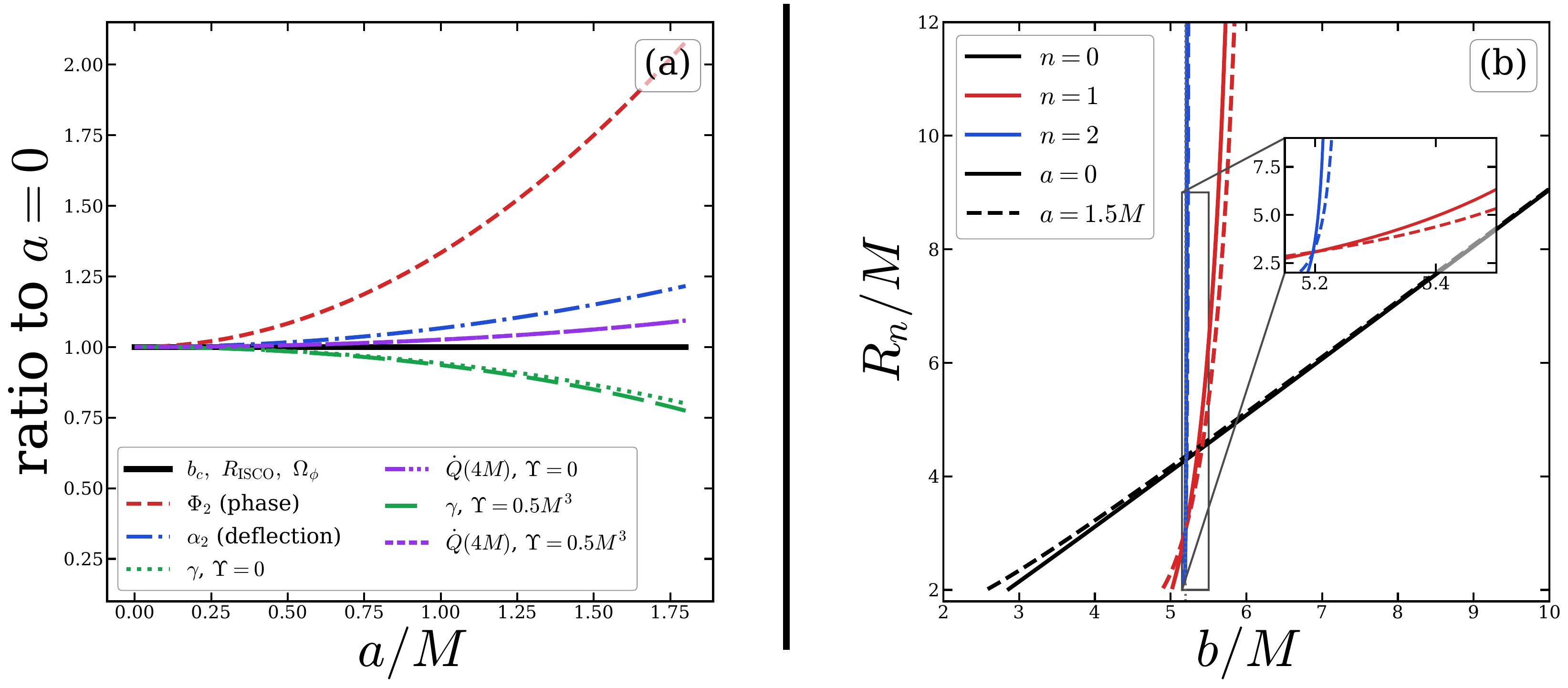}
\caption{(a) Ratios to the $a=0$ values of throat-blind quantities (solid black, the same curve for $b_c$, $R_{\rm ISCO}$ and $\Omega_\phi$), of the second-order neutrino phase and deflection coefficients, of the subring exponent and of the annihilation rate at $R_\nu=4M$, for $\Upsilon=0$ and $\Upsilon=0.5M^3$. (b) Emission radius of the $n$-th crossing of the equatorial plane versus impact parameter for a face-on observer in the Schwarzschild metric and in the hairless bounce with $a=1.5M$. The inset magnifies the $n=1$ and $n=2$ bands near the critical impact parameter, marked by the dotted vertical line.}
\label{fig:split}
\end{figure*}

Figure~\ref{fig:split}(a) summarizes the split as a function of the throat radius. The critical impact parameter, the ISCO and the Keplerian frequency stay at their $a=0$ values, the two second-order weak-field coefficients grow at different rates, the subring exponent decreases according to Eq.~\eqref{eq:gamma}, and the annihilation rate, discussed in Sec.~\ref{isec6}, rises. Panel (b) shows the emission radius of the $n$-th equatorial crossing for a face-on observer: the direct image barely moves, while the $n=1$ and $n=2$ lensing bands widen, as the inset makes clear near $b_c$.

\begin{table}[t]
\caption{Geometric data of the fiducial SV-KR geometries with $\lambda=2/3$ and of the charged-type case $\lambda=1$, in units of $M$. The last two columns compare the measured ratio of the widths of the $n=3$ and $n=2$ lensing bands with $e^{-\gamma}$ from Eq.~\eqref{eq:gamma}.}
\label{tab:geo}
\begin{ruledtabular}
\begin{tabular}{cccccccc}
$a$ & $\Upsilon$ & $R_+$ & $b_c$ & $R_{\rm ISCO}$ & $\gamma$ & $w_3/w_2$ & $e^{-\gamma}$\\
\hline
0 & 0 & 2.00000 & 5.19615 & 6.00000 & 3.14159 & 0.04274 & 0.04321\\
0.8 & 0 & 2.00000 & 5.19615 & 6.00000 & 3.02783 & 0.04785 & 0.04842\\
1.5 & 0 & 2.00000 & 5.19615 & 6.00000 & 2.72070 & 0.06488 & 0.06583\\
1.75 & 0 & 2.00000 & 5.19615 & 6.00000 & 2.55170 & 0.07680 & 0.07795\\
0 & 0.5 & 1.85464 & 5.03934 & 5.71764 & 2.96628 & 0.05084 & 0.05149\\
1.5 & 0.5 & 1.85464 & 5.03934 & 5.71764 & 2.52071 & 0.07915 & 0.08040\\
1.75 & 0.5 & 1.85464 & 5.03934 & 5.71764 & 2.33905 & 0.09540 & 0.09642\\
\hline
\multicolumn{8}{c}{$\lambda=1$}\\
1.5 & 0.5 & 1.70711 & 4.70960 & 5.16646 & 2.37959 & & 0.09262\\
\end{tabular}
\end{ruledtabular}
\end{table}

Table~\ref{tab:geo} also tests Eq.~\eqref{eq:gamma} independently. The widths $w_n$ of the lensing bands were obtained by integrating the orbit equation in the variable $u=1/R$ and locating the impact parameters at which a ray reaches the $n$-th crossing exactly at the horizon or exactly at infinity. Their ratio converges to $e^{-\gamma}$ as $n$ grows, and at $n=3$ it already agrees with the closed form to within about one part in seventy for every geometry. The Schwarzschild row reproduces $R_+=2M$, $b_c=3\sqrt3M$, $R_{\rm ISCO}=6M$ and $\gamma=\pi$, and the $\lambda=1$ row reproduces the charged black bounce with $Q^2=\Upsilon$, for which $R_+=(1+1/\sqrt2)M$ and $R_{\rm ph}=(3+\sqrt5)M/2$.

\section{Polarized images and photon subrings}\label{isec4}

The images of an optically and geometrically thin equatorial disk provide the cleanest test of the split, since the direct emission comes from throat-blind circular orbits while the lensed emission travels along throat-sensitive paths. This section describes the ray-tracing scheme, the treatment of polarization and the resulting images.

\subsection{Rays, emitters and polarization transport}

Null geodesics in a spherically symmetric spacetime are planar. In the orbital plane we integrate
\begin{equation}
\Big(\frac{du}{d\psi}\Big)^2=(1-a^2u^2)\Big[\frac{1}{b^2}-u^2f(1/u)\Big],\qquad u=\frac1R,\label{eq:orbit}
\end{equation}
in its second-order form, which is regular through turning points and starts exactly at $u=0$ with $du/d\psi=1/b$. For an observer at inclination $i$ and a pixel at screen radius $b$ and polar angle $\varphi$, the swept angle at the $n$-th crossing of the equator is $\psi_n=\psi_0+n\pi$ with $\tan\psi_0=-\cot i/\sin\varphi$ and $\psi_0\in(0,\pi)$. A table of $R(b,\psi)$ on a grid refined logarithmically toward $b_c$ maps every pixel to its emission radii.

Outside the ISCO the gas follows Keplerian orbits, Eq.~\eqref{eq:circ}, and inside it plunges with the ISCO energy and angular momentum. In the static frame the plunging velocity components are
\begin{equation}
v^{\hat\phi}=\frac{L_{\rm ISCO}\sqrt f}{RE_{\rm ISCO}},\qquad v^{\hat R}=-\sqrt{1-\frac{f}{E_{\rm ISCO}^2}\Big(1+\frac{L_{\rm ISCO}^2}{R^2}\Big)},\label{eq:plunge}
\end{equation}
which again depend only on $f(R)$ and $R$, so the entire emitter model is throat-blind and any change in the image comes from the light paths. The redshift is $g=\sqrt f/[\Gamma(1-\bm v\cdot\hat{\bm k})]$, with $\hat{\bm k}$ the photon direction in the static frame. Following Refs.~\cite{Narayan2021,Himwich2020}, the observed intensity is
\begin{equation}
I=g^{3+\alpha_\nu}J(R)\,|\hat{\bm k}'\times\hat{\bm B}'|^{1+\alpha_\nu}\,l_p,\qquad J=\frac{e^{-(R/22M)^4}}{R^2},\label{eq:emis}
\end{equation}
with spectral index $\alpha_\nu=1$, the path length $l_p=k'^{t}/|k'^{z}|$ capped at four, and primed quantities in the fluid frame. The local polarization is $\bm f'\propto\hat{\bm k}'\times\hat{\bm B}'$.

Real accretion flows are not smooth. Magnetized turbulence, sheared by differential rotation, organizes the emission into thin trailing filaments, which dominate both simulated and observed images. To represent this in the displayed images we multiply $J$ by a stochastic texture of unit mean,
\begin{align}
\mathcal{T}(R,\phi)&=\exp\Big[\sigma\,\mathcal{N}(R,\phi)-\frac{\sigma^2}{2}\Big],\nonumber\\
\mathcal{N}(R,\phi)&\propto\sum_{j}A_j\cos\big(Q_j\ln R+k_j\phi+\varphi_j\big),\label{eq:texture}
\end{align}
normalized to unit variance, with $260$ modes of random phase, $Q_j$ between $14$ and $130$, $k_j$ within two units of $Q_j\tan\vartheta$ for a pitch angle $\vartheta=0.16$, amplitudes $A_j\propto Q_j^{-0.35}$ and contrast $\sigma=1.1$. Since all wavevectors are nearly parallel in the $(\ln R,\phi)$ plane, the superposition forms long logarithmic-spiral filaments. The texture is a property of the emitter and is the same function of $(R,\phi)$ in every geometry, so it cannot mimic a throat effect. The polarimetric coefficients quoted below and in Table~\ref{tab:beta} refer to the smooth emissivity; with the texture, $\arg\beta_2$ of the full image at $i=17^\circ$ shifts by about $1^\circ$ and $|\beta_2|$ by less than one hundredth, while the throat-induced shift between the Schwarzschild and SV-KR images stays at $1.1^\circ$.

Transporting the polarization is simple in spherical symmetry. Let $\hat{\bm h}$ be the unit normal of the orbital plane and $\hat{\bm w}=\hat{\bm k}\times\hat{\bm h}$ the in-plane transverse direction in the static frame. Reflection symmetry about the orbital plane keeps $f_\perp=\bm f\cdot\hat{\bm h}$ constant under parallel transport, the norm $\bm f\cdot\bm f$ is constant, and both $f_\perp$ and $f_\parallel=\bm f\cdot\hat{\bm w}$ are unchanged by the gauge shift $\bm f\to\bm f+c\,\bm k$. Hence $f_\parallel$ is conserved as well, and the polarization on the screen is
\begin{equation}
\bm E_{\rm screen}=f_\perp\,\hat{\bm e}_\varphi-f_\parallel\,\hat{\bm e}_b,\label{eq:pol}
\end{equation}
where $\hat{\bm e}_b$ and $\hat{\bm e}_\varphi$ are the radial and azimuthal unit vectors of the screen. This replaces the Walker-Penrose constant used for Kerr \cite{Himwich2020} and is exact for any static spherically symmetric metric. As a check, a static face-on disk threaded by a radial field gives a purely azimuthal electric pattern with $\beta_2=-1$, and a toroidal field gives $\beta_2=+1$. We quantify the pattern by the coefficient \cite{Palumbo2020}
\begin{equation}
\beta_2=\frac{\int d^2x\,(Q+iU)\,e^{-2i\varphi_{\rm sky}}}{\int d^2x\,I},\label{eq:beta2}
\end{equation}
with $\varphi_{\rm sky}$ measured east of north, and by the net fractional polarization $m_{\rm net}=|\int(Q+iU)|/\int I$.

\subsection{Images}

\begin{figure*}[t]
\centering
\includegraphics[width=\textwidth]{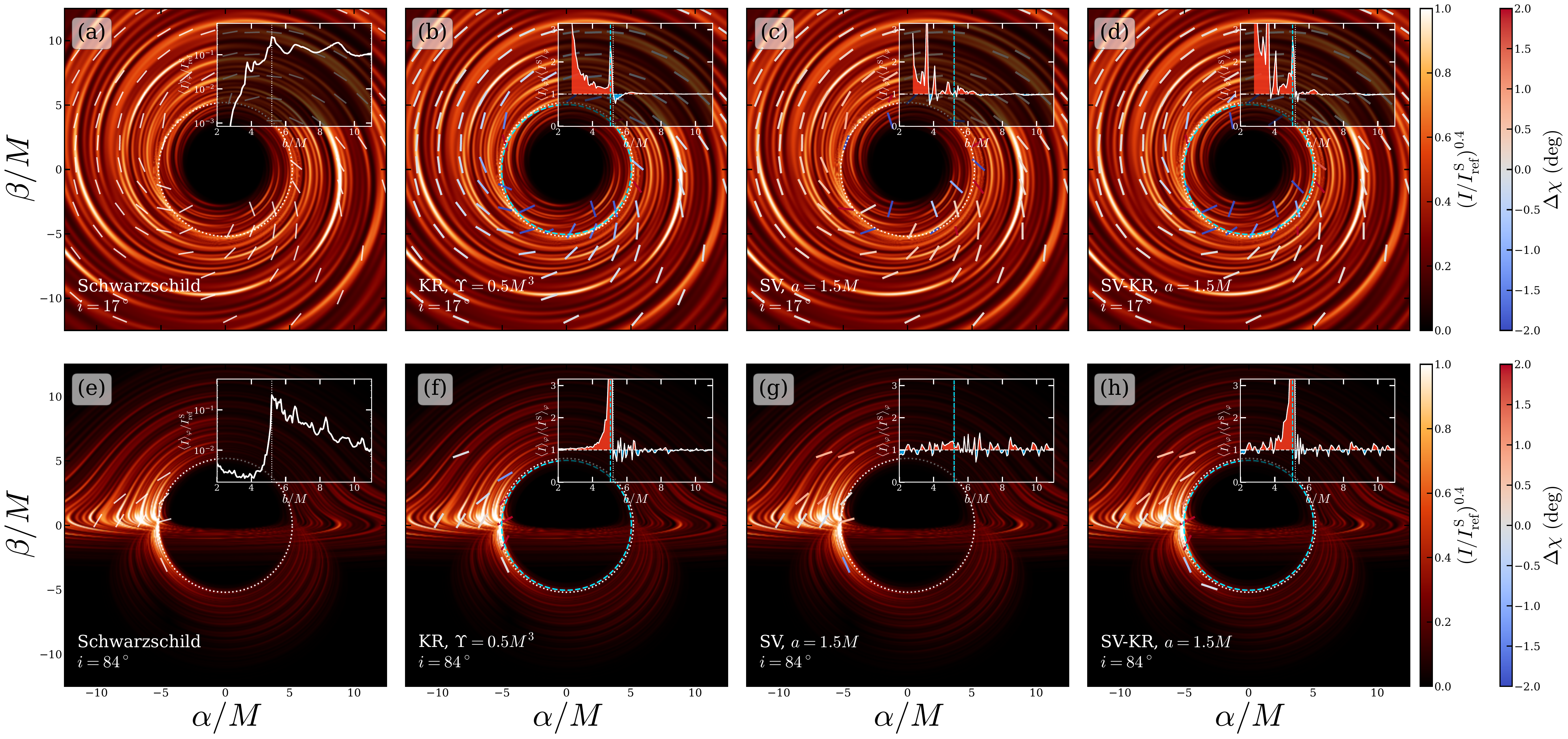}
\caption{Polarized images of a thin, turbulent Keplerian disk at $i=17^\circ$ (top row) and $i=84^\circ$ (bottom row) for the Schwarzschild metric, the Kalb-Ramond hair with $\Upsilon=0.5M^3$, the hairless bounce with $a=1.5M$ and the SV-KR geometry with both. All panels of a row share one color scale, $(I/I^{\rm S}_{\rm ref})^{0.4}$, with $I^{\rm S}_{\rm ref}$ the 99.6th percentile of the Schwarzschild intensity, so that brightness differences between geometries are real. Ticks show the electric vector position angle with length proportional to the fractional polarization; outside the Schwarzschild column their color gives the rotation $\Delta\chi$ relative to the Schwarzschild image at the same pixel. The dotted white circle is the Schwarzschild critical curve and the dashed cyan circle the critical curve of the geometry shown. Insets show the azimuthally averaged intensity $\langle I\rangle_\varphi$ versus screen radius for Schwarzschild and, in the other columns, its ratio to Schwarzschild, with excess in red and deficit in blue; dotted and dashed vertical lines mark the two critical impact parameters. The hair moves the ring inward and piles intensity up just inside the Schwarzschild critical curve, while the throat raises the intensity between $3M$ and $5M$ through its wider lensing bands.}
\label{fig:pol}
\end{figure*}

Figure~\ref{fig:pol} shows the four fiducial geometries at $i=17^\circ$, appropriate for M87*, and at $i=84^\circ$, for a fluid-frame field $\hat{\bm B}'\propto0.87\,\hat{\bm R}+0.5\,\hat{\bm z}$. The four columns look nearly identical, which is the visual form of the throat-blind statement: the critical curve is the same circle for the two hairless and for the two hairy geometries, and the direct image of the disk, which dominates the flux, comes from throat-blind emitters seen along paths that barely approach the throat. The polarization ticks spiral in the same sense in every panel. Near edge-on, the lensed image of the far side of the disk arches over the shadow and its underside appears below it, framed by the thin photon ring, while Doppler beaming makes the approaching side of the filaments far brighter than the receding side.

\begin{figure*}[t]
\centering
\includegraphics[width=\textwidth]{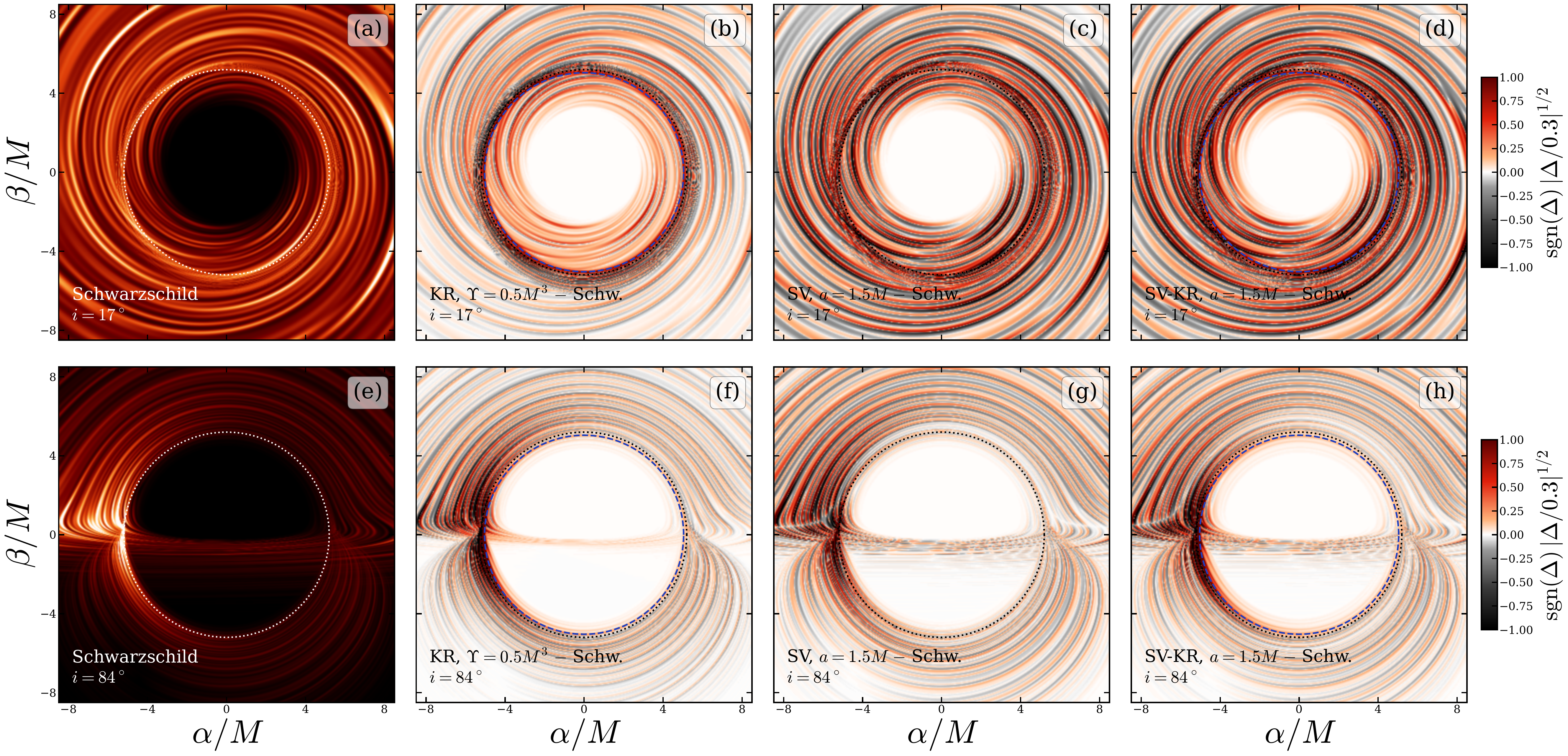}
\caption{Residual maps of the images of Fig.~\ref{fig:pol} relative to the Schwarzschild image, which is repeated in the first column, in a field of view of $\pm8.5M$. Color shows ${\rm sgn}(\Delta)|\Delta/0.3|^{1/2}$ with $\Delta=(I-I^{\rm S})/I^{\rm S}_{\rm ref}$, a stretch that keeps weak residuals visible next to strong ones, at $i=17^\circ$ (top) and $i=84^\circ$ (bottom). Red marks an excess and graphite gray to black a deficit of intensity. The dotted circle is the Schwarzschild critical curve and the dashed blue circle that of the geometry shown; the residuals of the hair cluster on the critical curve, those of the throat spread over the lensed filaments.}
\label{fig:blur}
\end{figure*}

The similarity of the four columns hides real differences, which Fig.~\ref{fig:blur} makes explicit by subtracting the Schwarzschild image pixel by pixel. The hair alone produces a residual concentrated on the critical curve, a positive and a negative ring that record the shrinking of $b_c$ from $5.196M$ to $5.039M$. The throat alone leaves the critical curve in place but displaces the lensed filaments, since their paths are governed by Eq.~\eqref{eq:orbit}: the residual appears as a sequence of interleaved red and dark arcs throughout the inner disk and in the lensed image of the far side. With both ingredients the two patterns add. Measured by the summed absolute residual normalized to the summed Schwarzschild intensity, the difference at $i=17^\circ$ is $0.07$ for the hair, $0.19$ for the throat and $0.24$ for both, and at $i=84^\circ$ it is $0.45$, $0.47$ and $0.57$, respectively. The fine structure of the disk is what makes the throat visible in these maps: a smooth emitter would hide most of the displacement, which is why flux-weighted quantities such as $\beta_2$ respond only weakly.

Blurring with a circular Gaussian of full width $20\,\mu$as, which corresponds to $5.3M$ for M87*, removes the photon ring as a separate feature and erases these residuals. The central brightness depression survives in all four geometries and its size tracks the hair rather than the throat. The blurring lowers $|\beta_2|$ from $0.870$ to $0.749$ in the Schwarzschild case and from $0.858$ to $0.734$ in the SV-KR case, while the phase changes by $2.5^\circ$ and $2.8^\circ$, respectively, so the blurred polarimetric pattern carries almost no trace of the throat.

\begin{figure*}[t]
\centering
\includegraphics[width=\textwidth]{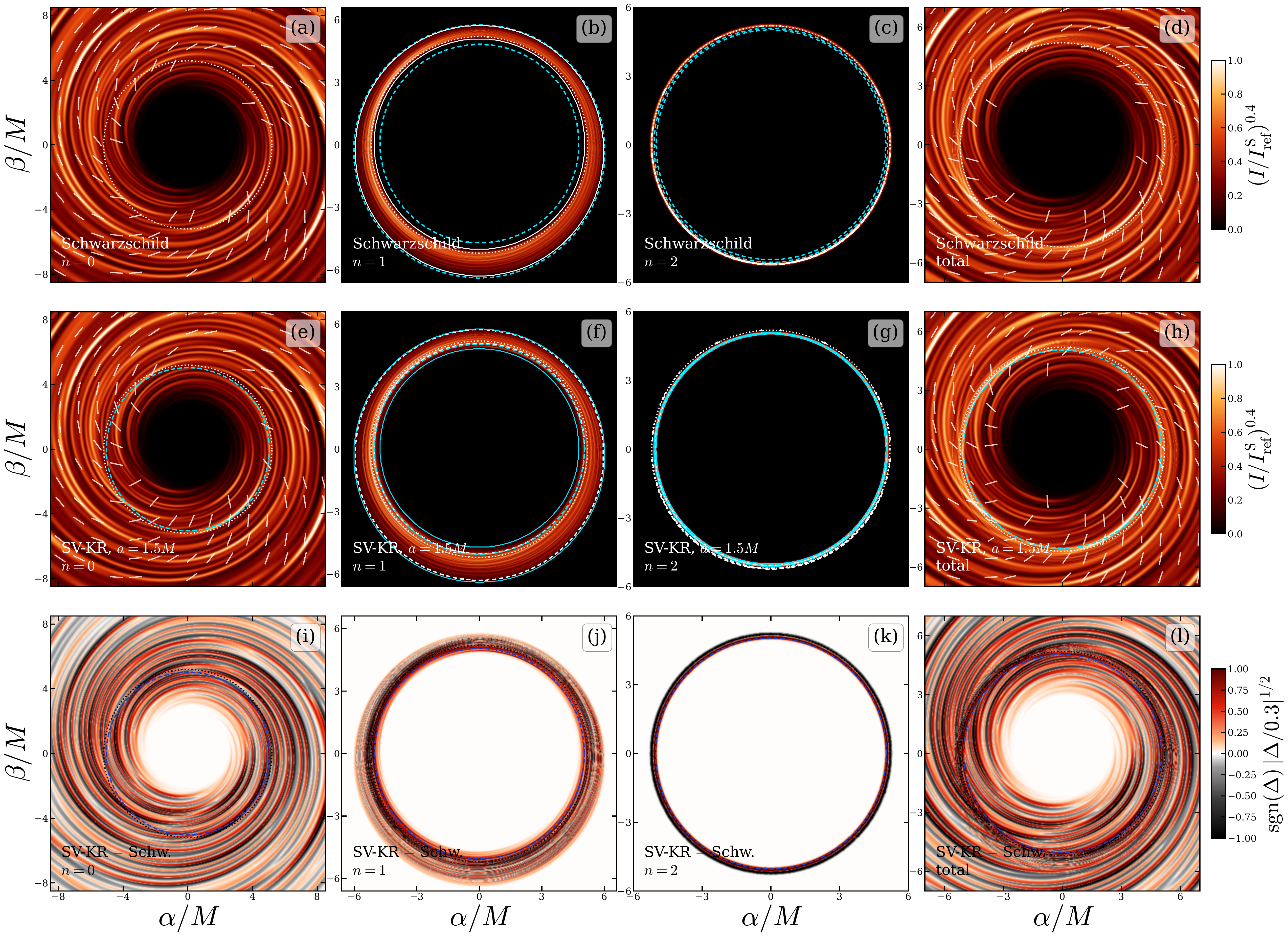}
\caption{Decomposition of the image at $i=17^\circ$ into the direct emission $n=0$, the first and second subrings $n=1$ and $n=2$, and their sum for the Schwarzschild metric (top) and the SV-KR geometry with $a=1.5M$, $\Upsilon=0.5M^3$ (bottom). The emissivity carries the texture of Eq.~\eqref{eq:texture}, and each column uses the color scale of the Schwarzschild panel above it, $(I/I^{\rm S}_{\rm ref})^{0.4}$, so the brighter SV-KR subrings are a physical effect. In the $n=1$ and $n=2$ panels the solid line traces the edge of the band of the geometry shown and the dashed line the band of the other geometry: the SV-KR band extends further inward. The bottom row shows the residual SV-KR minus Schwarzschild for each order with the stretch of Fig.~\ref{fig:blur}: the direct image shows displaced filaments, the $n=1$ residual an inner excess ring from the wider band, and the $n=2$ residual a bright pair of rings at the shifted critical curve. The subring panels use a narrower field of view; the $n=2$ ring, thinner than a pixel, is lightly smoothed and drawn without ticks.}
\label{fig:rings}
\end{figure*}

The throat becomes visible when the image is decomposed by the number of equatorial crossings. Figure~\ref{fig:rings} separates the $n=0$, $n=1$ and $n=2$ contributions for the Schwarzschild and the SV-KR geometries. The direct images differ little, but the $n=1$ band of the SV-KR case is broader on its inner side and both subrings are brighter. The integrated $n=1$ flux grows by a factor $1.39$ relative to Schwarzschild at $i=17^\circ$ and the $n=2$ flux by roughly a factor of two, consistent with the larger $e^{-\gamma}$ of Table~\ref{tab:geo}. Because the critical curve is fixed by the hair alone, a measured subring width at a given ring diameter isolates the throat through Eq.~\eqref{eq:gamma}.

\begin{figure*}[t]
\centering
\includegraphics[width=0.94\textwidth]{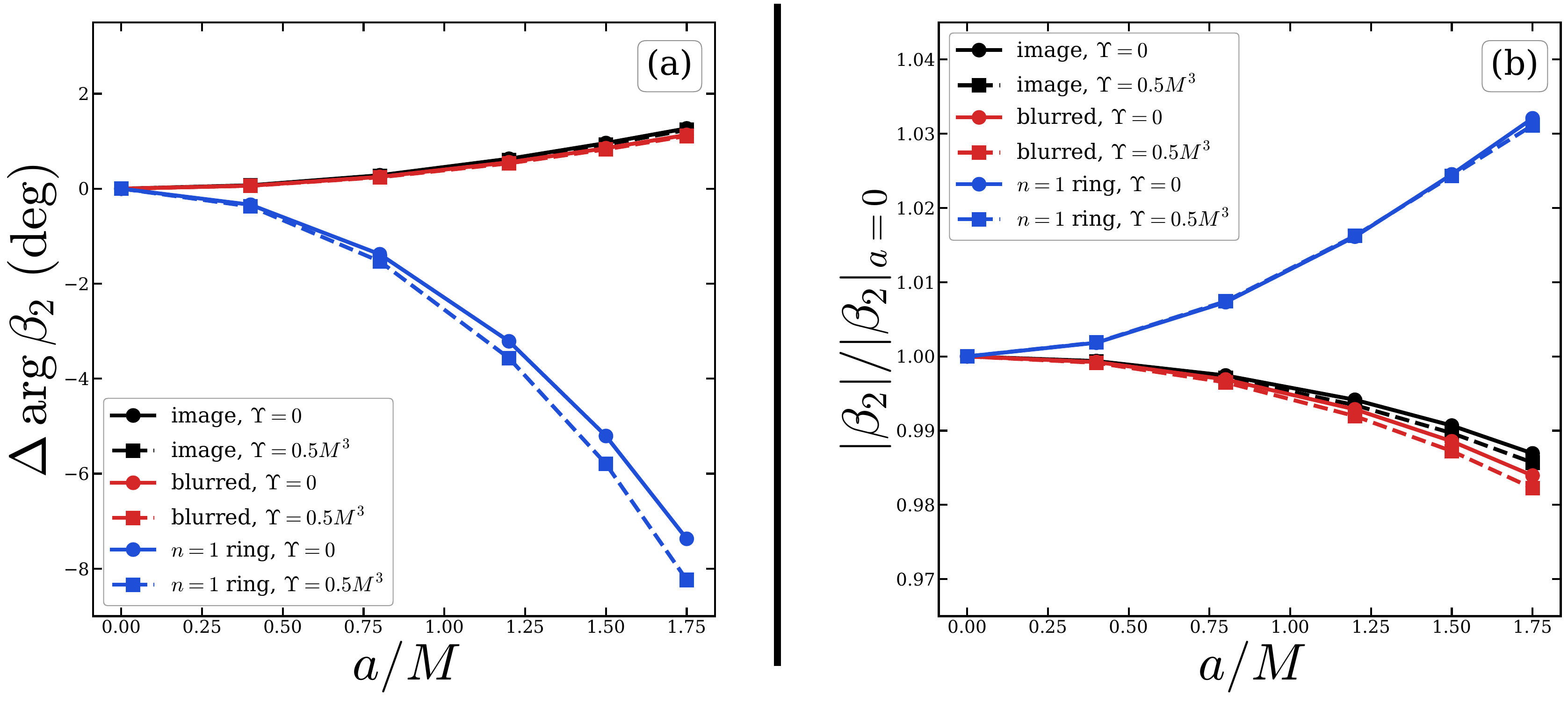}
\caption{(a) Shift of the phase of $\beta_2$ relative to $a=0$ and (b) modulus of $\beta_2$ relative to $a=0$, as functions of the throat radius at $i=17^\circ$ for the full image, the blurred image and the $n=1$ ring, for $\Upsilon=0$ (circles, solid) and $\Upsilon=0.5M^3$ (squares, dashed).}
\label{fig:beta}
\end{figure*}

The polarimetric counterpart is shown in Fig.~\ref{fig:beta}. Over the range $0\le a\le1.75M$ the phase of $\beta_2$ of the full image moves by about $1.3^\circ$ and its modulus by about one part in seventy, both for the unblurred and the blurred images. The $n=1$ ring alone responds roughly five times more strongly: its phase moves by $7.4^\circ$ for $\Upsilon=0$ and by $8.2^\circ$ for $\Upsilon=0.5M^3$, and its modulus grows by three parts in a hundred. Table~\ref{tab:beta} extends the comparison to purely radial, vertical and toroidal fields. The toroidal field is nearly insensitive to the throat. The radial field shifts the $n=1$ phase by $2.6^\circ$ and the mixed field by $5.0^\circ$ at $a=1.5M$. The vertical field moves the phase of the full image by about $3^\circ$, which is the largest response of the direct emission, because its polarization depends most strongly on the direction of the photon at the source. For $\lambda=1$ the full-image phase at $a=1.5M$, $\Upsilon=0.5M^2$ is $-134.67^\circ$, close to the $\lambda=2/3$ value, so the sensitivity of the polarization to the hair exponent is weak at fixed critical-curve size.

\begin{table}[t]
\caption{Polarimetric coefficient $\beta_2$ at $i=17^\circ$ for four fluid-frame field geometries, for the Schwarzschild metric (S), the hairless bounce with $a=1.5M$ (SV) and the SV-KR geometry with $a=1.5M$, $\Upsilon=0.5M^3$. Phases are in degrees; $m_{\rm net}$ refers to the full image.}
\label{tab:beta}
\begin{ruledtabular}
\begin{tabular}{llccccc}
Field & Model & $|\beta_2|$ & $\arg\beta_2$ & $|\beta_2^{n=1}|$ & $\arg\beta_2^{n=1}$ & $m_{\rm net}$\\
\hline
mixed & S & 0.870 & $-135.19$ & 0.677 & $-98.97$ & 0.0423\\
 & SV & 0.862 & $-134.22$ & 0.693 & $-104.17$ & 0.0414\\
 & SV-KR & 0.858 & $-134.02$ & 0.693 & $-103.94$ & 0.0412\\
radial & S & 0.961 & $-163.63$ & 0.914 & $-141.83$ & 0.0234\\
 & SV & 0.958 & $-162.96$ & 0.909 & $-144.46$ & 0.0234\\
 & SV-KR & 0.956 & $-162.82$ & 0.906 & $-144.44$ & 0.0235\\
vertical & S & 0.673 & 89.72 & 0.644 & 143.81 & 0.1121\\
 & SV & 0.670 & 92.47 & 0.609 & 142.30 & 0.1087\\
 & SV-KR & 0.666 & 92.80 & 0.601 & 141.64 & 0.1068\\
toroidal & S & 0.989 & 2.45 & 0.954 & 5.82 & 0.0200\\
 & SV & 0.988 & 2.57 & 0.956 & 5.96 & 0.0208\\
 & SV-KR & 0.988 & 2.63 & 0.956 & 5.97 & 0.0207\\
\end{tabular}
\end{ruledtabular}
\end{table}

\section{Neutrino phases and multipath flavor oscillations}\label{isec5}

Neutrinos produced near a compact object probe the radial integral of Eq.~\eqref{eq:I} directly, and they do so in a way that photons cannot, because the oscillation phase depends on the neutrino masses. This section derives the phase in the SV-KR geometry, checks its weak-field expansion and studies the interference of paths for a source close to the bounce.

\subsection{Oscillation phase}

A flavor state $|\nu_\alpha\rangle=\sum_kU^*_{\alpha k}|\nu_k\rangle$ propagating from the source S to the detector D accumulates for each mass eigenstate the phase $\Phi_k=\int_S^D p_\mu^{(k)}dx^\mu$. Evaluated along the light ray and expanded to first order in $m_k^2/E^2$, it becomes \cite{Fornengo1997,Cardall1997,Visinelli2015}
\begin{align}
\Phi_k&=\frac{m_k^2}{2E}\int_S^D\frac{\sqrt{-g_{tt}g_{rr}}\,dr}{\sqrt{1-b^2f/R^2}}\nonumber\\
&=\frac{m_k^2}{2E}\int_S^D\frac{dR}{\sqrt{h}\sqrt{1-b^2f/R^2}},\label{eq:phase}
\end{align}
where $E$ is the energy at infinity, and the second form uses $-g_{tt}g_{rr}=1$ in the coordinate $r$ of Eq.~\eqref{eq:svkr}. The radial measure of the phase is exactly the factor $h^{-1/2}$ that enters the orbit equation. For radial propagation in the Schwarzschild metric the phase reduces to $m_k^2\Delta r/(2E)$ \cite{Fornengo1997}. In the SV-KR geometry radial propagation gives $m_k^2\Delta r/(2E)$ in the coordinate $r$, which corresponds to the proper areal path $\Delta\sqrt{R^2-a^2}$, so that a source at areal radius $R_S$ acquires the extra phase $m_k^2(R_S-\sqrt{R_S^2-a^2})/(2E)$ relative to the hairless and throatless case. Away from the radial direction the phase of one leg, regularized at infinity, is $m_k^2\mathcal{P}(b)/(2E)$ with $\mathcal{P}$ of Eq.~\eqref{eq:Pexp}.

Figure~\ref{fig:nu}(a) compares the exact quadrature of $\mathcal{P}$ with the expansion of Eq.~\eqref{eq:Pexp}. The quantity $b(\mathcal{P}-M)/M^2$ approaches $3\pi/4$ for the hairless geometries without throat and $3\pi/4+\pi a^2/(4M^2)$ with throat, confirming the coefficient of $a^2$. At $b=3000M$ the exact values are $2.3588$ and $4.1267$ against the asymptotic $2.3562$ and $4.1233$, and the residual falls as $1/b$. The hair with $n=3$ enters only at order $b^{-2}$ and is invisible on this scale, which is the weak-field version of the hair independence of Eq.~\eqref{eq:invert}. The quadrature itself was checked against an independent arbitrary-precision evaluation and agrees to ten digits.

\subsection{Multipath flavor probabilities}

In the weak-lensing regime studied in Refs.~\cite{Swami2020,Chakrabarty2022,Alloqulov2025}, where the impact parameters are of the order of the Einstein radius, Eq.~\eqref{eq:Pexp} shows that the throat changes the phase difference between paths by $\pi a^2(b_1^{-1}-b_2^{-1})m_k^2/(4E)$. This is of second order in $M/b$ and negligible for any realistic Einstein radius. The throat becomes accessible when the neutrinos are emitted close to the compact object, as in the inner region of a hyperaccreting flow or the neutrinosphere of a collapsar. We therefore place a point source at areal radius $R_S=10M$ on the negative $x$ axis and a detector at infinity in the direction $\varphi_D$. A ray leaving the source with local emission angle $\theta$, measured from the outward radial direction, has $b=R_S\sin\theta/\sqrt{f(R_S)}$. It escapes either directly or after a turning point, provided $b>b_c$ in the second case, and it reaches the detector when its total swept angle equals $\pi\mp\varphi_D$ modulo $2\pi$. For each such path $p$ we compute $\psi$ and the regularized radial integral $\mathcal{I}_p$, and assign the amplitude $\sqrt{w_p}$ with
\begin{equation}
w_p=\left|\frac{\sin\theta_p}{\sin\varphi_D}\frac{d\theta}{d\psi}\right|_p ,\label{eq:weight}
\end{equation}
the ratio of emitted to received solid angle. Earlier studies assigned equal amplitudes to all paths, which overstates the contribution of the paths that wind around the photon sphere. The flavor transition probability is then
\begin{equation}
P_{\alpha\beta}=\frac{\left|\sum_kU_{\beta k}U^*_{\alpha k}\sum_p\sqrt{w_p}\,e^{-im_k^2\mathcal{I}_p/2E}\right|^2}{\sum_k|U_{\alpha k}|^2\left|\sum_p\sqrt{w_p}\,e^{-im_k^2\mathcal{I}_p/2E}\right|^2},\label{eq:prob}
\end{equation}
which is normalized by construction. We use the mixing angles $\theta_{12}=33.41^\circ$, $\theta_{13}=8.58^\circ$, $\theta_{23}=49.1^\circ$, $\delta_{\rm CP}=0$ and the splittings $\Delta m^2_{21}=7.41\times10^{-5}\,{\rm eV}^2$, $|\Delta m^2_{3\ell}|=2.51\times10^{-3}\,{\rm eV}^2$ close to the global fit \cite{Esteban2020}, a lightest mass of $0.02\,$eV, a lens of ten solar masses ($M=14.77\,$km) and $E=10\,$MeV.

\begin{figure*}[t]
\centering
\includegraphics[width=\textwidth]{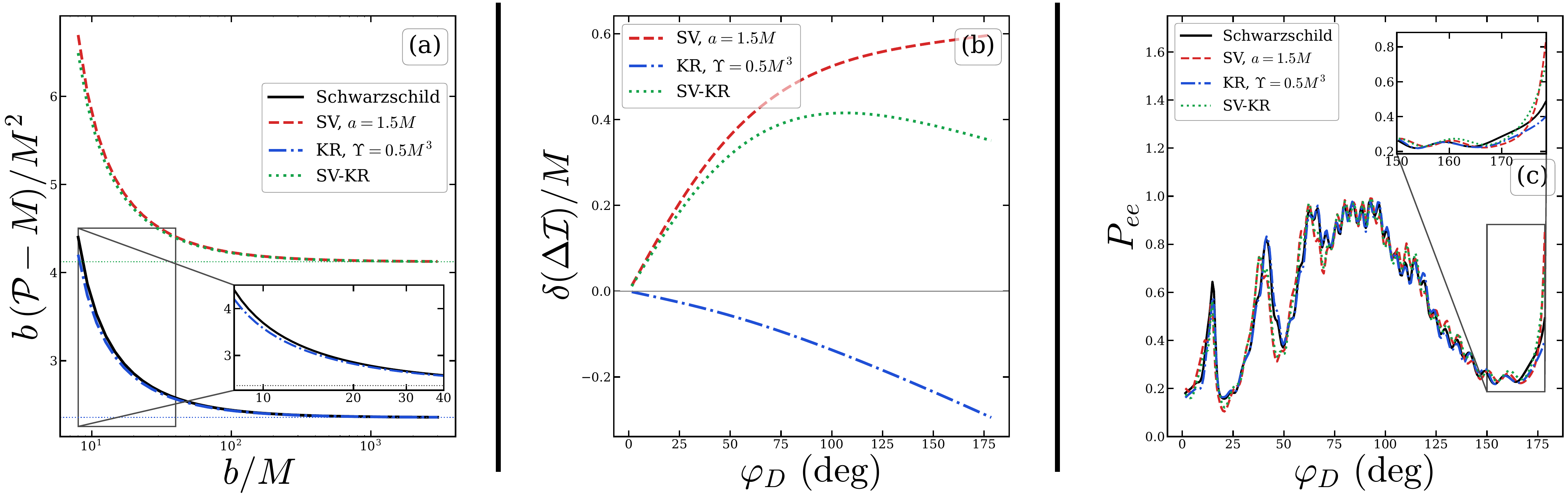}
\caption{(a) Exact second-order coefficient $b(\mathcal{P}-M)/M^2$ of the one-leg neutrino phase integral versus impact parameter; dotted lines show the asymptotic values $3\pi/4$ and $3\pi/4+\pi a^2/(4M^2)$ of Eq.~\eqref{eq:Pexp}, and the inset magnifies the range $8M\le b\le40M$, where the hair lowers the coefficient of the throatless geometries. (b) Change of the radial-integral difference of the two primary paths relative to Schwarzschild for a source at $R_S=10M$, as a function of the detector direction. (c) Electron-neutrino survival probability for the normal ordering, a ten-solar-mass lens and $E=10\,$MeV; the inset magnifies the region $150^\circ\le\varphi_D\le178.5^\circ$.}
\label{fig:nu}
\end{figure*}

Figure~\ref{fig:nu}(b) shows the change of the path difference between the two primary paths, $\delta(\Delta\mathcal{I})=\Delta\mathcal{I}-\Delta\mathcal{I}_{\rm Schw}$. The throat lengthens the difference monotonically as the detector moves toward the source side, reaching $0.60M$ for $a=1.5M$ near $\varphi_D=180^\circ$, while the hair shortens it by up to $0.29M$. For the ten-solar-mass lens the throat contribution is $8.8\,$km, which for $\Delta m^2_{31}$ at $10\,$MeV corresponds to a phase shift of $5.6\,$rad. With both ingredients the shift is $0.35M$ and the hair partly compensates the throat. At $\varphi_D=90^\circ$ the four paths with $k\le1$ have weights $0.80$, $0.027$, $8.7\times10^{-4}$ and $3.7\times10^{-5}$ in the Schwarzschild metric; in the SV-KR geometry the third and fourth paths are about twice and three times as bright, reflecting the smaller $\gamma$.

The resulting electron-neutrino survival probability is shown in Fig.~\ref{fig:nu}(c) for the normal ordering. The fast oscillation in $\varphi_D$ is driven by the large path difference and by $\Delta m^2_{31}$, and the slower envelope by $\Delta m^2_{21}$. The throat shifts the fast pattern in phase, with differences of a few tenths in $P_{ee}$ near $\varphi_D=180^\circ$, where the paths are most unequal. At $\varphi_D=90^\circ$ the survival probability is $0.866$ in the Schwarzschild metric and $0.918$ in the SV-KR geometry for the normal ordering, and $0.879$ and $0.904$ for the inverted ordering. The ordering matters at the same level as the throat: at fixed geometry the two orderings differ by about one hundredth in $P_{ee}$ at this angle, while the throat changes $P_{ee}$ by five hundredths for the normal and by two and a half hundredths for the inverted ordering.

\section{Pair-annihilation energy deposition}\label{isec6}

The third messenger is the energy deposited by $\nu\bar\nu\to e^+e^-$ above a neutrino-emitting surface. Here the throat enters through the proper volume, which makes this observable throat-sensitive at leading order rather than through path differences.

We model the emitter as an isothermal neutrinosphere of areal radius $R_\nu$, above the photon sphere, with luminosity $L_\infty$ fixed at infinity. The local temperature and the angle subtended by the neutrinosphere depend only on $f$ and $R$, $T(R)=T(R_\nu)\sqrt{f(R_\nu)/f(R)}$ and $\sin\theta_R=(R_\nu/R)\sqrt{f(R)/f(R_\nu)}$, and the local deposition rate is proportional to $T^9G(\cos\theta_R)$ with $G(x)=(1-x)^4(x^2+4x+5)$ \cite{Asano2000,Lambiase2020}. The rate measured at infinity integrates this over the proper volume weighted by the redshift, $\sqrt{-g}\,d^3x=R^2h^{-1/2}dR\,d\Omega$. Normalizing to the Newtonian rate at the same $L_\infty$ we obtain
\begin{equation}
\frac{\dot Q}{\dot Q_{\rm N}}=f_\nu^{-9/4}\frac{\int_1^\infty dy\,y^2\big(f_\nu/f\big)^{9/2}G(x)\,h^{-1/2}}{\int_1^\infty dy\,y^2\,G\big(\sqrt{1-y^{-2}}\big)},\label{eq:dep}
\end{equation}
with $y=R/R_\nu$, $f_\nu=f(R_\nu)$ and $x=\cos\theta_R$. The factor $f_\nu^{-9/4}$ converts $L_\infty$ to the local temperature, and the only place where the throat enters is $h^{-1/2}$ in the numerator. For small $a$, the relative change of the rate is $\delta\dot Q/\dot Q\simeq\frac{a^2}{2}\langle R^{-2}\rangle$, with the average taken over the deposition profile.

The small-$a$ estimate is accurate. At $R_\nu=4M$ in the hairless case the exact relative enhancement is $2.270\times10^{-3}$ for $a=0.3M$ and $2.617\times10^{-2}$ for $a=M$, against $2.262\times10^{-3}$ and $2.513\times10^{-2}$ from the profile average; at $R_\nu=6M$ the corresponding pairs are $1.004\times10^{-3}$ against $1.002\times10^{-3}$, and $1.133\times10^{-2}$ against $1.113\times10^{-2}$. The deposition is concentrated near the emitter: about $0.99$ of the total rate is deposited within $2R_\nu$ for both radii, which explains why the throat effect follows the surface value of $a^2/(2R^2)$ closely.

\begin{figure*}[t]
\centering
\includegraphics[width=0.94\textwidth]{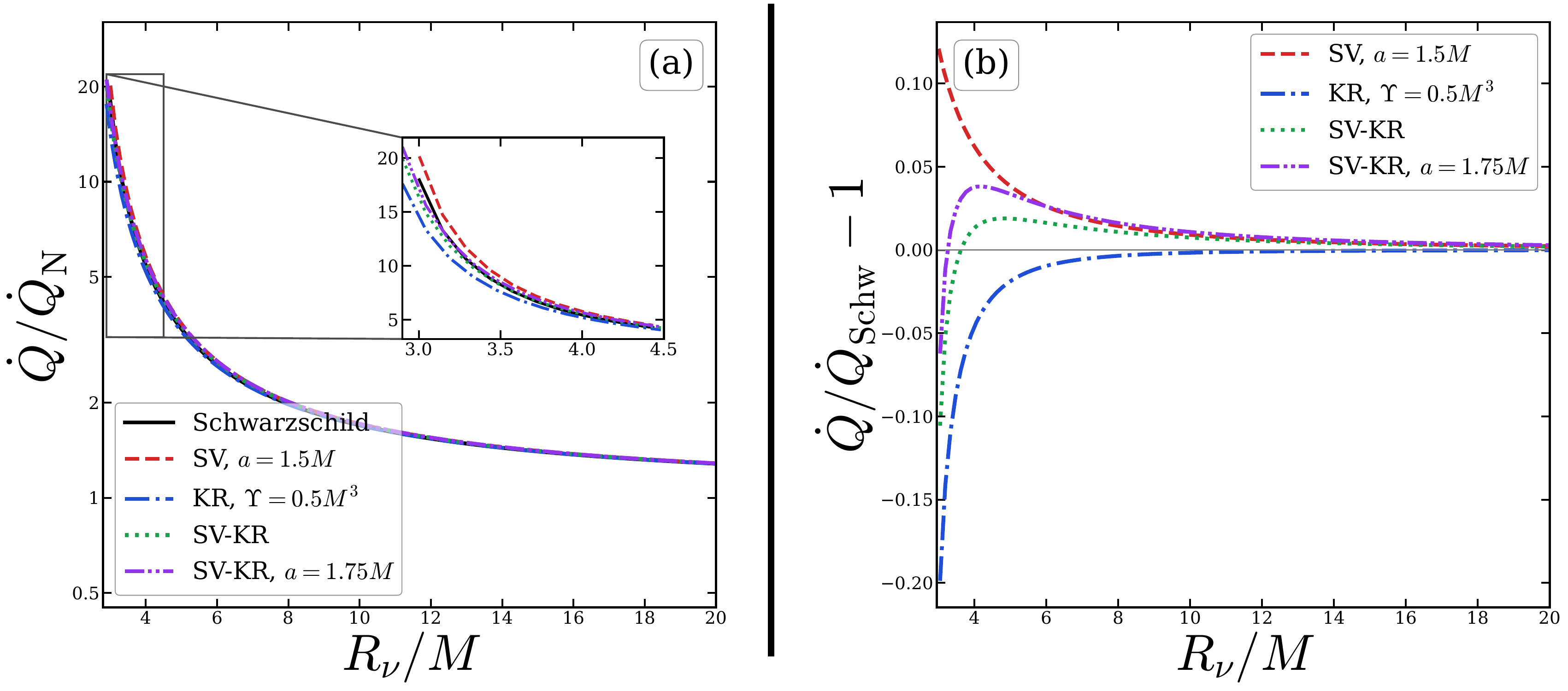}
\caption{(a) Pair-annihilation energy deposition rate normalized to the Newtonian rate at fixed luminosity at infinity, as a function of the neutrinosphere radius; the inset magnifies the region near the photon sphere. (b) Relative change of the rate with respect to the Schwarzschild metric.}
\label{fig:dep}
\end{figure*}

Figure~\ref{fig:dep}(a) shows the ratio \eqref{eq:dep}. At large $R_\nu$ it tends to unity, as it must, with the value $1.049$ at $R_\nu=100M$ in the Schwarzschild metric. Close to the photon sphere gravity amplifies the deposition strongly: at $R_\nu=4M$ the ratio is $5.443$ for Schwarzschild, $5.781$ for the hairless bounce with $a=1.5M$, $5.204$ for the hair alone and $5.527$ for the SV-KR geometry. The throat therefore raises the rate by $0.062$ in relative terms at fixed hair, slightly below the surface value $a^2/(2R_\nu^2)=0.070$ of the small-$a$ estimate because the deposition extends beyond the neutrinosphere. Panel (b) shows that this relative enhancement decays as $R_\nu^{-2}$, and that the hair with $\Upsilon>0$ acts in the opposite direction, so that in the SV-KR geometry the two effects cancel near $R_\nu\simeq3.5M$ for $a=1.5M$ and the throat wins farther out. At $R_\nu=3M$ the SV-KR rate is $16.13$ for $a=1.5M$ and $16.95$ for $a=1.75M$, against $14.40$ for the hair alone. The independent evaluation of Eq.~\eqref{eq:dep} with a different quadrature reproduces the SV-KR value at $R_\nu=3M$ to fourteen digits.

Two remarks put these numbers in context. The enhancement over the Newtonian rate near the photon sphere is driven by $f_\nu^{-9/4}$ and by the factor $(f_\nu/f)^{9/2}$, which describe the gravitational blueshift of the neutrinos falling back toward the emitter; these factors are throat-blind and set the overall scale, as for the modified-gravity backgrounds of Refs.~\cite{Lambiase2020,Lambiase2022,AraujoFilho2024SV}. The throat adds a geometric factor on top of them, which is why it appears as a relative correction of order $a^2/R_\nu^2$ that is the same for any metric function. For a black-hole accretion torus rather than a spherical neutrinosphere the angular integration changes \cite{Asano2001,Birkl2007}, but the proper-volume factor, and hence the structure of the throat correction, does not.

\section{Observational windows}\label{isec7}

The split of Sec.~\ref{isec3} dictates which data can constrain which parameter. The size of the critical curve constrains only the hair. Writing $\delta=b_c/(3\sqrt3M)-1$, the fiducial hair with $n=3$ gives $\delta=-0.030$ for $\Upsilon=0.5M^3$ and $\delta=-0.084$ at extremality, while negative hair increases the ring and reaches $\delta=+0.1$ only at $\Upsilon=-2.30M^3$. An EHT window of one tenth in $\delta$, of the size reported for Sagittarius A* \cite{EHT2022}, therefore leaves every positive hair up to extremality allowed. For $\lambda=1$ the same window excludes $\Upsilon>0.53M^2$. In both cases the throat radius is completely unconstrained by the shadow, as is evident from Eq.~\eqref{eq:bc}, and comparisons of black-hole mimickers with the Sagittarius A* image \cite{Shaikh2023} cannot fix it either.

It is useful to quantify the size of the throat signals. For the hairless bounce the subring exponent decreases from $\pi$ to $2.72$ at $a=1.5M$, so the flux ratio of consecutive subrings rises from $0.043$ to $0.066$, a factor of $1.5$; at fixed $b_c$ this is a large change. For the hairy bounce with $\Upsilon=0.5M^3$ the same throat moves $e^{-\gamma}$ from $0.051$ to $0.080$. In polarimetry, the $n=1$ phase shift of $5^\circ$ at $a=1.5M$ is small compared with the spread between field geometries in Table~\ref{tab:beta}, which exceeds a hundred degrees, so a throat measurement through $\beta_2$ needs independent information on the field geometry; the full-image shift of about $1^\circ$ lies well below the uncertainties of present EHT polarization maps \cite{EHT2021}. For neutrinos from a ten-solar-mass engine the throat changes the primary path difference by up to $8.8\,$km, while the corresponding arrival-time difference between paths, of order $26M\simeq1.3\,$ms, sets the coherence time a wave packet would need to keep the interference terms of Eq.~\eqref{eq:prob}.

Three channels see the throat. The first is the ratio of successive subring fluxes or widths, which measures $\gamma$ and, once the hair is known from the ring diameter, gives $a$ through Eq.~\eqref{eq:gamma}. The universal interferometric signature of the photon ring \cite{Johnson2020} makes this an observable for long baselines, and Fig.~\ref{fig:geo}(b) shows that $\gamma/\pi$ ranges from about $0.67$ to $1.07$ over the black-bounce domain. The second is the phase of $\beta_2$ restricted to the $n=1$ ring, which moves by about five degrees at $a\simeq1.5M$, roughly five times more than the phase of the full image. The third is the neutrino channel, through either the multipath interference of Sec.~\ref{isec5} for neutrinos from the inner accretion region, or the deposition enhancement of Sec.~\ref{isec6}, or, for a lens observed in both photons and neutrinos, the inversion \eqref{eq:invert}.

\section{Conclusions}\label{isec8}

We have shown that the Simpson-Visser regularization of the Kalb-Ramond power-law hair leaves the metric function, written in the areal radius, exactly as it was and moves all dependence on the throat into the radial measure $h^{-1/2}=(1-a^2/R^2)^{-1/2}$. This single statement sorts the observables. The critical curve, the ISCO, Keplerian frequencies and local redshifts are throat-blind; the Lyapunov exponent, the subring demagnification, the radial epicyclic frequency, light deflection, the neutrino oscillation phase and the pair-annihilation deposition are throat-sensitive. We derived the exact law $\gamma(a)=\gamma(0)\sqrt{1-a^2/R_{\rm ph}^2}$, the weak-field coefficients of deflection and phase, including the hair coefficients for any fall-off exponent, and the combination of photon and neutrino data that separates mass and throat. The effective source has a null-energy violation set by the throat alone, $\rho+p_r=-2a^2f/(\kappa R^4)$.

The polarized images confirm the split. The direct image and the full polarimetric pattern barely respond to the throat, while the $n=1$ ring widens and its $\beta_2$ phase moves by about $7^\circ$ to $8^\circ$ over the black-bounce range. For neutrinos emitted at $10M$ from a stellar-mass lens, the throat changes the interfering path difference by up to $0.6M$ and the flavor survival pattern by order unity, and near the photon sphere it raises the annihilation deposition, an effect that the positive hair partly offsets.

All benchmarks were recovered: the Schwarzschild values $R_+=2M$, $b_c=3\sqrt3M$, $R_{\rm ISCO}=6M$, $\gamma=\pi$, the Simpson-Visser deflection coefficient, the Reissner-Nordstr\"om deflection coefficient for $n=2$, the charged black bounce for $\lambda=1$, the radial phase of Fornengo and collaborators, and the Newtonian limit of the deposition. The closed forms and an independent kernel evaluation agree to ten significant figures for the geometric quantities and the phase integral and to fourteen for the deposition, while the band-width ratios approach $e^{-\gamma}$ to two significant figures at $n=3$. Our neutrinos are test particles described by plane waves along geometric-optics paths, coherence between paths with different arrival times is assumed rather than derived, the disk is optically and geometrically thin, and only the static black-bounce branch is treated, so none of the results bears on ringdown or gravitational-wave signals.

In the near future, we plan to extend the analysis to the rotating Kalb-Ramond bounce, where the critical curve itself is no longer circular and the split between throat-blind and throat-sensitive quantities must be reformulated with the Carter constant. A wave-packet treatment of the multipath neutrino amplitudes, which fixes how much interference survives the time delays between paths, and a computation of the radial epicyclic frequencies for quasiperiodic oscillations are the other two tasks on our agenda.

\begin{acknowledgments}
F.A. acknowledges the Inter University Centre for Astronomy and Astrophysics (IUCAA), Pune, India, for granting a visiting associateship. \.{I}.S. is grateful to Eastern Mediterranean University (EMU), the Scientific and Technological Research Council of T\"urkiye (T\"UB\.{I}TAK), the Turkish Academic Network and Information Center (ANKOS) and the Sponsoring Consortium for Open Access Publishing in Particle Physics (SCOAP3) for their support. He also acknowledges the networking support of COST Actions CA18108, CA22113, CA21106, CA23130, CA23115, and CA21136.
\end{acknowledgments}

\section*{Author contributions}
F.A. conceived the study and carried out the geodesic, lensing and weak-field calculations. A.A. contributed to the analysis of the shadow, the deflection angle and the neutrino phases, and checked the calculations. \.{I}.S. carried out the ray-traced imaging, polarimetry and pair-annihilation analysis and prepared the figures. All authors discussed the results, wrote the manuscript via Overleaf, and approved its final version.

\section*{Data availability}
No observational data were generated in this study. The computational worksheets that produce the figures and tables are available from the corresponding author upon reasonable request.

\appendix

\section{Weak-field coefficients}\label{app:weak}

We sketch the derivation of Eqs.~\eqref{eq:defl}--\eqref{eq:cn}. Writing $x=R_0/R$ and $b^2=R_0^2/f(R_0)$, the integrals \eqref{eq:psi} and \eqref{eq:I} become
\begin{align}
\psi&=\frac{b}{R_0}\int_0^1\frac{dx}{\sqrt{1-a^2x^2/R_0^2}\,\sqrt{1-x^2F(x)}},\label{eq:A1}\\
\mathcal{P}&=R_0\int_0^1\frac{dx}{x^2}\bigg[\frac{1}{\sqrt{1-a^2x^2/R_0^2}\,\sqrt{1-x^2F(x)}}\nonumber\\
&\qquad\qquad-\frac{1}{\sqrt{1-x^2}}\bigg],\label{eq:A2}
\end{align}
with $F(x)=f(R_0/x)/f(R_0)$. The subtraction in Eq.~\eqref{eq:A2} removes the flat-space path $\sqrt{R_{\rm e}^2-R_0^2}$, and the difference between $\sqrt{R_{\rm e}^2-R_0^2}$ and $R_{\rm e}$ vanishes at infinity. Since $1-x^2F(x)$ vanishes linearly at $x=1$, the substitution $x=1-s^2$ makes both integrands regular. The throat contribution follows at once: to first order in $a^2$,
\begin{align}
\delta\psi&=\frac{a^2}{2R_0^2}\int_0^1\frac{x^2dx}{\sqrt{1-x^2}}=\frac{\pi a^2}{8b^2},\label{eq:A3}\\
\delta\mathcal{P}&=\frac{a^2}{2R_0}\int_0^1\frac{dx}{\sqrt{1-x^2}}=\frac{\pi a^2}{4b},\label{eq:A4}
\end{align}
which doubles to the $a^2$ term of $\alpha=2\psi-\pi$ in Eq.~\eqref{eq:defl} and gives the $a^2$ term of Eq.~\eqref{eq:Pexp}. For the hair, a term $\Upsilon R^{-n}$ in $f$ changes $1-x^2F(x)$ by $-\Upsilon R_0^{-n}x^2(x^n-1)$ to first order. Expanding the integrands and integrating with $\int_0^1x^{k}(1-x^2)^{-1/2}dx=\sqrt\pi\,\Gamma(\frac{k+1}{2})/[2\Gamma(\frac{k}{2}+1)]$ gives the coefficients of Eq.~\eqref{eq:cn}, and the ratio $c^\alpha_n/c^\Phi_n=2(n+1)/n$ comes out of the same expansion. The mass terms are obtained in the same way to third order in $M/R_0$, after re-expressing $R_0$ in terms of $b$.

We verified each coefficient against direct quadrature with thirty-digit arithmetic. For $n=2,3,4$ the numerical hair coefficients at $b=10^4M$ are $-0.7858$, $-1.0006$ and $-1.1800$, against the closed forms $-\pi/4$, $-1$ and $-3\pi/8$; the coefficient of $Ma^2/b^3$ in the deflection is $2.6668$ against $8/3$; and the Schwarzschild third-order coefficient approaches $128/3$.

\section{Numerical methods and convergence}\label{app:num}

The geometric quantities of Table~\ref{tab:geo} were obtained by root finding on Eqs.~\eqref{eq:bc}--\eqref{eq:isco} with a tolerance of $10^{-15}$ and checked against an independent symbolic and numerical kernel, which agreed to ten significant figures for $R_+$, $R_{\rm ph}$, $b_c$, $\lambda_L$, $\gamma$ and $R_{\rm ISCO}$ in the five geometries checked, including the $\lambda=1$ case. The band widths $w_n$ were computed by integrating Eq.~\eqref{eq:orbit} with an eighth-order Runge-Kutta scheme at relative tolerance $10^{-12}$ and bracketing the band edges to $10^{-15}$ in $b$.

The images use a table of $R(b,\psi)$ with about 1400 impact parameters, 520 of them distributed logarithmically within $0.6M$ of $b_c$ down to $10^{-9}M$, and 3600 values of $\psi$ up to $4.6\pi$, so that crossings up to $n=3$ are resolved. Pixel grids of $760^2$ points were averaged in blocks of four to suppress aliasing of the thin rings. For the SV-KR geometry at $i=17^\circ$, $\arg\beta_2$ changes from $-134.10^\circ$ to $-133.99^\circ$ and $-133.98^\circ$ for grids of $240^2$, $480^2$ and $720^2$ pixels, and $|\beta_2|$ stays at $0.8577$ to four digits, so the image coefficients quoted in the text are converged to about $0.1^\circ$ in phase and $10^{-3}$ in modulus.

The neutrino integrals were evaluated with the substitution $x=1-s^2$ of Appendix~\ref{app:weak} and adaptive quadrature. Against thirty-digit arithmetic the one-leg integral $\mathcal{P}$ for the SV-KR geometry agrees to $2\times10^{-12}$ at $b=20M$ and to $2\times10^{-11}$ at $b=200M$. For the multipath problem a fan of about 3000 emission angles, refined logarithmically toward the critical angle, was interpolated with monotone cubic splines in $\psi$, which gives the emission angle, the radial integral and the solid-angle weight of each path. The deposition ratio was integrated to $y=400$; raising the cutoff to $4000$ changes it by less than $10^{-12}$.

\bibliographystyle{apsrev4-2}
\bibliography{ref}

\end{document}